# Decentralized Control for Heterogeneous Battery Energy Storage System

Yusuke Hakuta*, Yasushi Amano*, Tomohiko Jimbo**, and Shuji Tomura*

\* *Toyota Central R&D Labs., Inc., 41-1, Yokomichi, Nagakute, Aichi 480-1192 Japan
(e-mail: yusuke-hakuta@mosk.tytlabs.co.jp).*

\** *Toyota Motor Corporation, 1, Toyota-cho, Toyota, Aichi 471-8571 Japan*

**Abstract**: Battery energy storage systems (BESSs) are essential for stable power supply in renewable energy systems that can operate in all weather. Future BESSs will be massive and pluggable with several heterogeneous batteries. In this paper, a novel decentralized control method for a heterogeneous BESS is proposed, in which each battery autonomously operates based on its characteristics. First, a control method that uses only one broadcast signal for each type of battery is proposed. Second, the asymptotic stability of the tracking error is proved. Third, numerical simulations confirm that the proposed control method has robust tracking performance of the total electric power to the demanded power when some batteries fail and are detached from the system. Last, in order to suppress degradation of battery, equalization of the state of charge is achieved for each type of battery without communication among the batteries.

## 1. INTRODUCTION

Renewable energy has been under development for more than a decade in many countries to help resolve environmental problems such as global warming and air pollution. However, since the amount of electric power generated by these energy sources is weather dependent, large-scale energy storage systems consisting of components such as batteries, photovoltaic cells, and fuel cells have been introduced to adjust for electricity supply and demand (Choi et al. (2012); Arani et al. (2019)). In particular, several studies on battery energy storage systems (BESSs) have been reported (Lawder et al. (2014); Takeda et al. (2014); Georgious et al. (2016); Xu et al. (2017); Tobajas et al. (2022); Iwafune et al. (2019)).

In most BESSs, the electric power of the batteries is controlled by a centralized controller, which can easily ensure sufficient optimality and stability. The centralized controller monitors factors such as capacity, temperature, and voltage limitations on its server, and computes the optimal output power for each battery. As the number of batteries increases, the operating cost of the server increases. Furthermore, these systems may be vulnerable because each power supply operation relies on the server. To address this problem, decentralized and distributed control methods have been proposed (Arani et al. (2019); Amano et al. (2021); Akutsu et al. (2017)).

We focus on a heterogeneous BESS in this paper. Using heterogeneous batteries has the advantage that each battery has its own characteristics and performance. In the future, massive and pluggable BESSs with several heterogeneous batteries will be practically implemented. In this paper, a novel decentralized control method is proposed in which each battery autonomously operates based on its characteristics. The contributions of this study are as follows:

- A proposed novel decentralized control method is presented in which heterogeneous batteries can operate independently to perform to their specific characteristics.

- Asymptotic stability of the tracking error of the total electric power to the demanded power is proved.

- Robust tracking performance when some batteries fail and are detached from the system is validated by numerical simulations.

- Equalization of the state of charge (SOC) for each type of battery without communication among the batteries is validated by a numerical simulation.

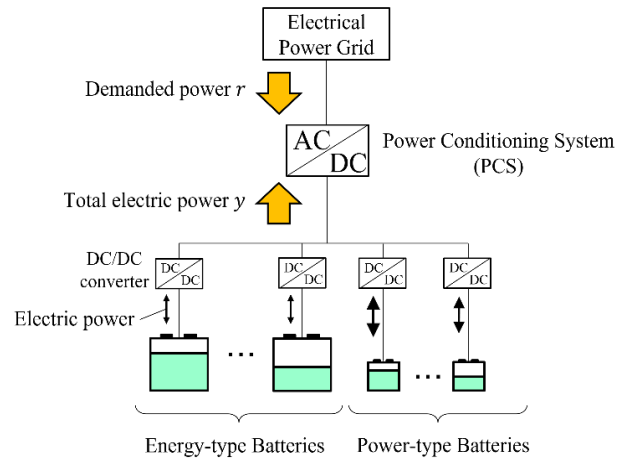

Fig. 1. Configuration of heterogenous BESS

The organization of this paper is as follows: In Section 2, the control problem addressed in this paper is formulated. Section 3 describes the proposed novel decentralized controller. In Section 4, the effectiveness of the proposed controller is demonstrated in numerical simulations. Section 5 provides the conclusions.

## 2. PROBLEM FORMULATION

Figure 1 shows the configuration of the heterogeneous battery energy storage system we analyzed. This system consists of high-capacity assembled batteries (energy-type batteries), high-power assembled batteries (power-type batteries), a power conditioning system (PCS), and DC/DC converters. The PCS converts the DC electric power discharged by the batteries into AC electric power. Each battery is equipped with a DC/DC converter. For simplicity, only batteries are considered as energy storage sources in this paper. The electric losses for the PCS and DC/DC converters are assumed to be zero.

We take the observable output $y$ to be the total electric power generated by all the batteries, and the reference $r$ to be the electric power demanded from the electrical power grid (demanded power). For $y > 0$, electric power is discharged from the batteries, and for $y < 0$ the batteries are charged. The electric power of each battery is the control input.

In this study, the formulation of the control problem requires the following:

- The closed-loop system should be stable.
- The total electric power $y$ should track the demanded power $r$.
- The SOC for batteries should be equalized to suppress battery degradation as much as possible.
- In a decentralized manner, the battery $i$ determines the electric power $u_i$ autonomously without communication among the batteries.
- Energy-type batteries operate more slowly than power-type batteries.

## 3. DECENTRALIZED CONTROL

We propose a novel decentralized control method in which heterogeneous batteries can behave differently to utilize their characteristics. The block diagram is shown in Fig. 2.

An energy-type battery $E_i (i = 1, \cdots, I)$ generates its own electric power $u_{E_i}$. It is determined using only one signal $e_E$ as follows:

$$\dot{\varphi}_{E_i} = K_{E_i} e_E$$
$$u_{E_i} = \sigma_{E_i}(\varphi_{E_i}), \qquad (1)$$

where $\varphi_{E_i}$ is the state of the controller, $K_{E_i}$ is the gain, and $\sigma_{E_i}$ is the switching function of the battery $E_i$. $e_E$ is the filtered error given by

$$T_f \frac{de_E}{dt} + e_E = r - y_E, \qquad (2)$$

where $r - y_E$ is broadcast from the electrical power grid, $T_f$ is a time constant, and $y_E = \sum_{i=1}^{I} u_{E_i}$. A first-order lag element with $T_f$ enables the energy-type batteries to operate slowly.

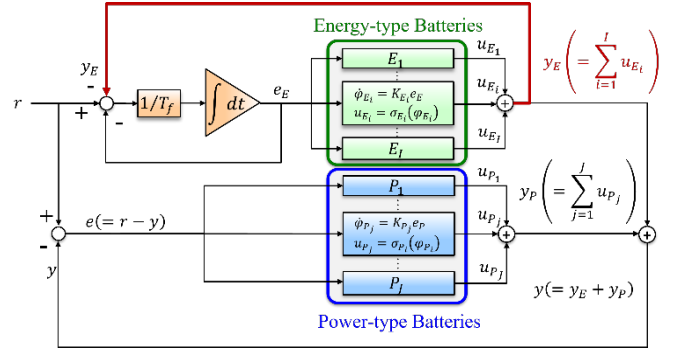

Fig. 2. Block diagram of proposed control method

A power-type battery $P_j (j = 1, \cdots, J)$ generates its own electric power $u_{P_j}$. It is also determined using only one signal $e (= r - y)$ as follows:

$$\dot{\varphi}_{P_j} = K_{P_j} e$$
$$u_{P_j} = \sigma_{P_j}(\varphi_{P_j}). \qquad (3)$$

Here, $\varphi_{P_j}$ is the state of the controller, $K_{P_j}$ is the gain, and $\sigma_{P_j}$ is the switching function of the battery $P_j$.

Figure 3 (a) shows the variable gain $K_k$ with upper bounds $K_k^D$ and $K_k^C$. $K_k$ depends on the SOC $s_k$ for the battery $k$ required to equalize the SOC of all batteries. Here, for the energy-type batteries and for the power-type batteries, $k = E_i (i = 1, \cdots, I)$ and $k = P_j (j = 1, \cdots, J)$, respectively. $s_{max}$ and $s_{min}$ are the maximum and minimum SOC values, respectively. For $k = E_i (i = 1, \cdots, I)$, we chose $r_{E_i} = r_E$, where $r_E$ represents a steady state with the demanded power $r$, and is defined in (B.2) in Appendix B. For $k = E_i (i = 1, \cdots, I)$, we chose $r_{P_j} = r$.

Figure 3 (b) shows the switching function $\sigma_k$. In the discharge demand from the electrical power grid, $u_k$ is zero for $\varphi_k \leq 0$ and is the maximal discharge power $\overline{U}_k^D$ for $\varphi_k \geq \overline{\varphi}_k^D$. In the charge demand, $u_k$ is zero for $\varphi_k \geq 0$ and the maximal charge power $\underline{U}_k^C$ for $\varphi_k \leq \underline{\varphi}_k^C$.

Based on (1), (2), and (3), energy-type batteries mainly operate in a steady state and power-type batteries operate in a transient state. Furthermore, for batteries governed by (1), (2), and (3), the following theorem holds.

**Theorem 1.** Consider the error system based on (1), (2) and (3) shown in Fig. 2. The tracking error $e$ and $e_E$ satisfy $e \to 0$ and $e_E \to 0$ as time $t \to \infty$, respectively.

**Proof.** See Appendix A.

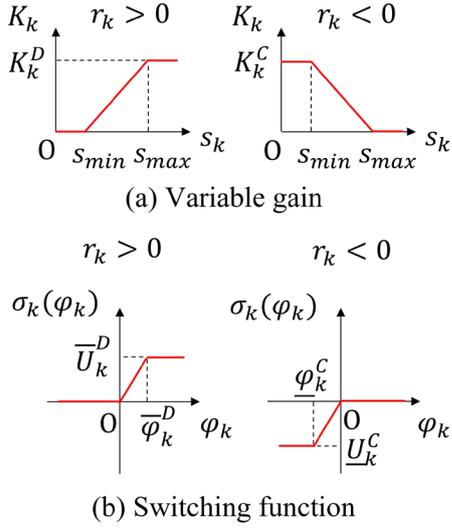

(a) Variable gain

(b) Switching function

Fig. 3. Controller for battery $k$

## 4. NUMERICAL SIMULATION

The proposed control method was simulated under the conditions given in Section 4.1. Section 4.2 briefly explains the configuration of a conventional centralized control method. In Section 4.3, the results of the simulation are shown.

### 4.1 Conditions

Five energy-type batteries and five power-type batteries were used, namely $I = J = 5$. Tables 1 and 2 show the specifications of the energy-type and power-type batteries, respectively. OCV stands for open circuit voltage.

$\overline{\varphi}_{E_i}^D$ and $\overline{\varphi}_{P_j}^D$ were 1, and $\underline{\varphi}_{E_i}^C$ and $\underline{\varphi}_{P_j}^C$ were $-1$ for all $i$ $(i = 1, \cdots, I)$ and $j$ $(j = 1, \cdots, J)$, respectively. $K_{E_i}^D$ and $K_{E_i}^C$ were fixed at $3 \times 10^{-7}$ for all $i$. $K_{P_j}^D$ and $K_{P_j}^C$ were fixed at $1 \times 10^{-6}$ for all $j$. The time step for the simulation was 0.01 seconds. The control period was 0.1 seconds and the time-delay for control was 0.3 seconds. The time constant $T_f$ was

Table 1. Specifications of energy-type batteries

| $i$ | OCV [V] | Internal resistance [Ω] | Capacity [Ah] | Initial SOC | $s_{max}$ | $s_{min}$ | $\overline{U}_{E_i}^D$ [kW] | $\underline{U}_{E_i}^C$ [kW] |
|---|---|---|---|---|---|---|---|---|
| 1 | 80 | 0.1 | 15 | 0.7 | 0.8 | 0.2 | 0.75 | -0.75 |
| 2 | | | | 0.6 | | | | |
| 3 | | | | 0.5 | | | | |
| 4 | | | | 0.4 | | | | |
| 5 | | | | 0.3 | | | | |

Table 2. Specifications of power-type batteries

| $j$ | OCV [V] | Internal resistance [Ω] | Capacity [Ah] | Initial SOC | $s_{max}$ | $s_{min}$ | $\overline{U}_{P_j}^D$ [kW] | $\underline{U}_{P_j}^C$ [kW] |
|---|---|---|---|---|---|---|---|---|
| 1 | 80 | 0.5 | 4 | 0.7 | 0.8 | 0.2 | 3 | -3 |
| 2 | | | | 0.65 | | | | |
| 3 | | | | 0.6 | | | | |
| 4 | | | | 0.55 | | | | |
| 5 | | | | 0.5 | | | | |

10 seconds. One minute after the start, the demanded power $r$ was expressed as a square wave, which has two processes, a 20-minute discharge ($r = 3\text{kW}$) and 20-minute charge ($r = -3\text{kW}$).

We performed simulations using the programming language MATLAB R2012b developed by MathWorks.

### 4.2 Conventional centralized control method

The centralized control method introduced by Takeda et al. (2014) was employed in this study for comparison with the proposed control method. The demanded power $r$ is separated into two parts, power in a steady state $r_E$ and power in a transient state $r - r_E$. Here, $r_E$ is obtained by applying a first-order lag element to $r$. The centralized controller receives the SOC for all batteries and determines the corresponding reference electric power for each battery. See Appendix B for details.

### 4.3 Results

We evaluated the effectiveness of the proposed control method from the following three viewpoints for a comparison with the conventional centralized control method.

1. Tracking error of the total electric power $y$ to the demanded power $r$.

2. Robust tracking performance of the total electric power $y$ to the demanded power $r$ when some batteries failed and were then detached from the system.

3. Equalization of the SOC for each type of battery

Figures 4(a) and 4(b) show the simulation results for the conventional centralized control method and the proposed control method, respectively, demonstrating the trackability of $y$ to $r$. In the conventional centralized control method, $y$ tracked $r$. However, the reference electric power for the energy-type batteries 1 and 2 exceeded the maximum output power in the time range between 1.5 minutes and 21 minutes. This means that these batteries cannot generate their reference electric power due to the constraint of maximum output power. The other batteries did not cover the shortage of electric power generated by energy-type batteries 1 and 2. As a result, $y$ did not reach $r$ in this time range. On the other hand, in the proposed control method, the energy-type batteries were operated in a steady state and the power-type batteries were operated during transient states of $r$ with their electric power converging to zero in the steady state. $y$ tracked $r$ as well as the conventional centralized control method.

Next, to consider the robustness of the tracking against battery failure, energy-type batteries 2 and 4 were assumed to fail at 11 minutes and 31 minutes, respectively. After that, the two batteries were detached from the system. In the conventional centralized control method, the centralized controller noticed the battery failure 5 minutes after failure by monitoring the electric power for each battery regularly. The centralized controller continued to provide the two batteries with their reference electric power based on the SOC

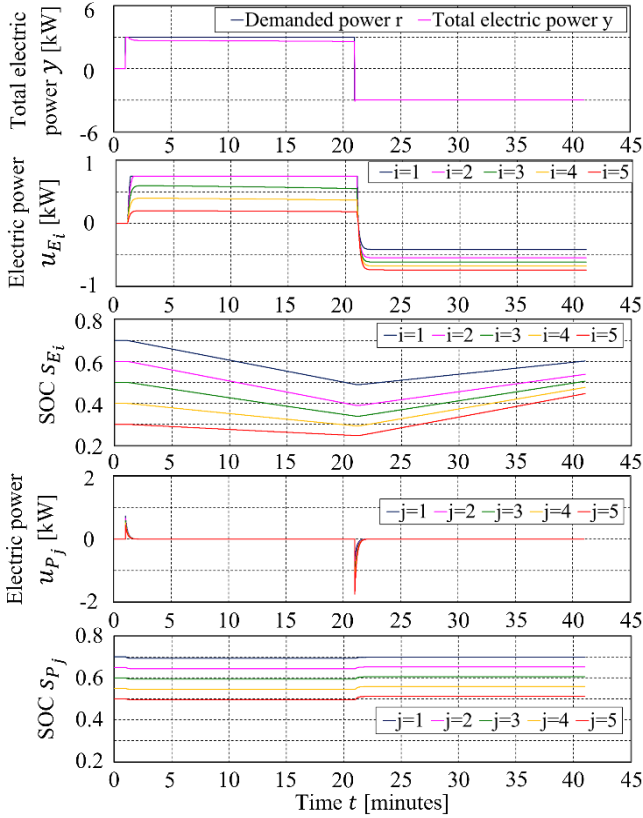
(a) Conventional centralized control method

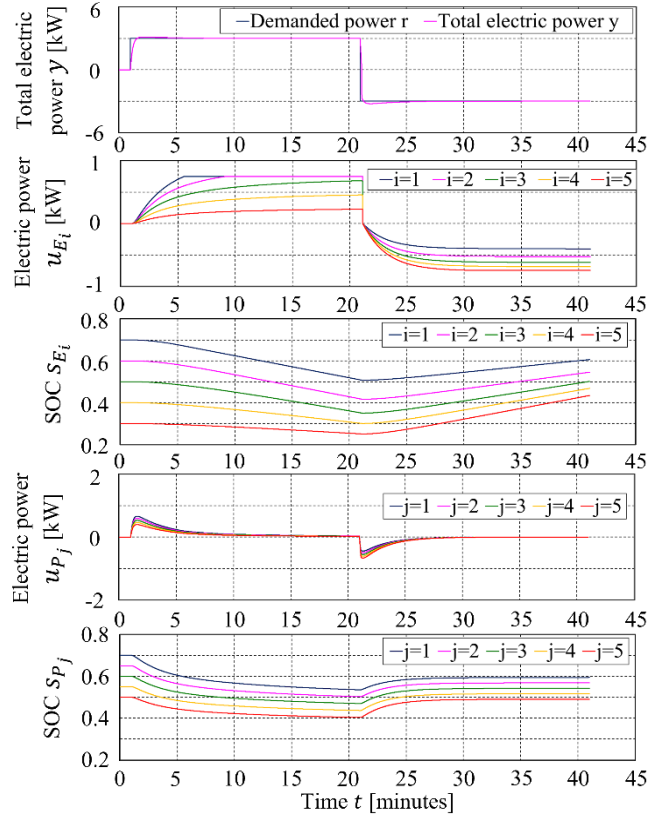
(b) Proposed control method

Fig. 4. Battery electric power and SOC for both methods

received just before their failure. Figures 5(a) and 5(b) show the simulation results for the conventional centralized control method and the proposed control method under the above conditions, respectively. In the conventional centralized control method, the other batteries did not cover the shortage of electric power generated by energy-type batteries 2 and 4. Therefore, $y$ did not track $r$. In contrast, the other batteries covered the power shortage in the proposed control method, and $y$ tracked $r$.

Finally, we investigated equalization of the SOC for each type of battery. Figures 6(a) and 6(b) show the simulation results for the energy-type batteries and the power-type batteries in the proposed control method, where the demanded power is four cycles of the square waves described in Section 4.1. The SOCs for both types of battery were equalized. This result indicates that it is effective to use the variable gain shown in Fig. 3 for achieving equalization of the SOC without communication among the batteries.

## 6. CONCLUSIONS

In this study, a novel decentralized control method was proposed for appropriately operating energy storage systems with several heterogeneous batteries. The proposed method has a simple configuration where only one broadcast signal is used for each type of battery, and the asymptotic stability of the tracking error of the total electric power to the demanded power was proved. In addition, numerical simulations confirmed that the proposed control method exhibits robust tracking of the total electric power to the demanded power when some batteries fail and are detached from the system. Furthermore, the variable gain enables us to achieve equalization of the SOCs for both types of battery without communication among the batteries. In the future, the proposed control method will be useful for pluggable large-scale BESSs.

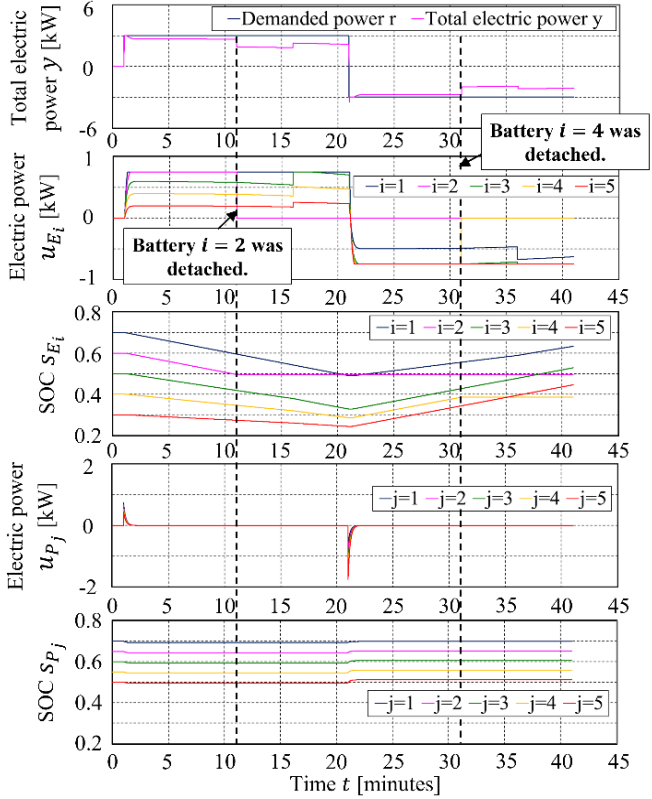
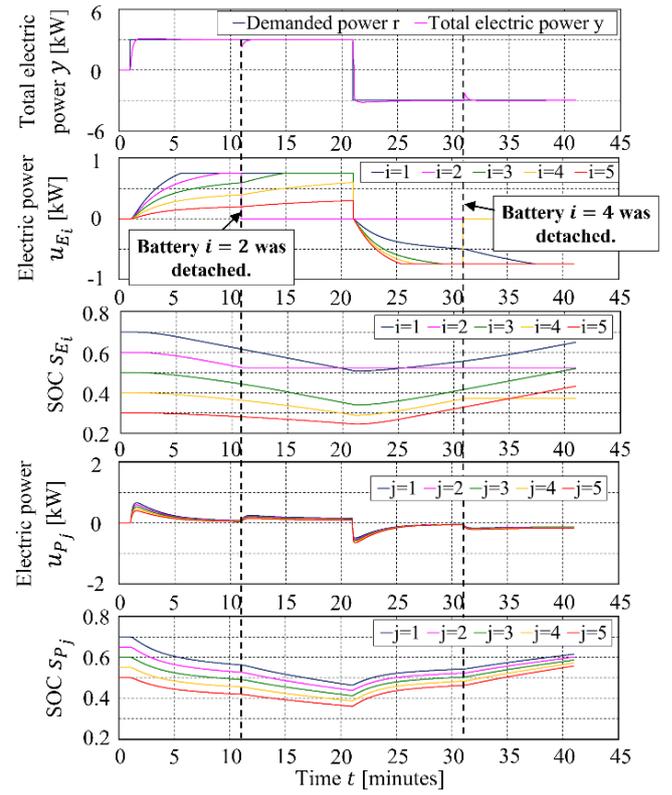

(a) Conventional centralized control method

(b) Proposed control method

Fig. 5. Battery electric power and SOC when batteries $i = 2, 4$ were detached from the system

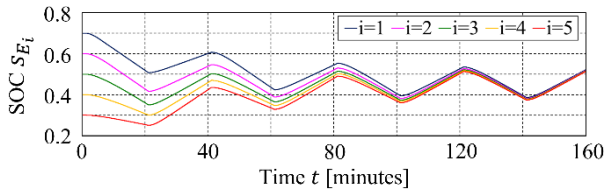
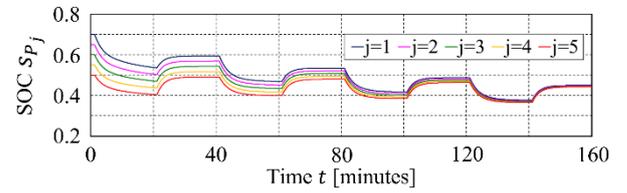

(a) Energy-type batteries

(b) Power-type batteries

Fig. 6. Battery SOC for the proposed control method (demanded power is 4 cycles of square waves)

Plants with Storage. 2017 11th *Asian Control Conference (ASCC)*.

## Appendix A. PROOF FOR THEOREM 1

First, we prove that $e_E \to 0$ as time $t \to \infty$. We introduce a function $\tilde{V}$ given by

$$\tilde{V} = \frac{e_E^2}{2} + \frac{1}{T_f}\sum_{i=1}^{I} V_{E_i}. \quad (A.1)$$

Here, $V_{E_i}$ is defined as follows:

$$V_{E_i} = \int_0^{\varphi_{E_i}} \frac{\sigma_{E_i}(\varphi_i) - r_i}{K_{E_i}} d\varphi_i \geq 0, \quad r = \sum_{i=1}^{I} r_i. \quad (A.2)$$

The proof for $V_{E_i} \geq 0$ is given in Amano et al. (2021). $e_E^2 \geq 0$ leads to $\tilde{V} \geq 0$. The derivative of the function $\tilde{V}$ with respect to $t$ is

$$\frac{d\tilde{V}}{dt} = e_E \frac{de_E}{dt} + \frac{1}{T_f}\sum_{i=1}^{I} \frac{dV_{E_i}}{dt}. \quad (A.3)$$

Each term on the right-hand of (A.3) can be expressed as follows (Amano et al. (2021)):

$$e_E \frac{de_E}{dt} = \frac{e_E}{T_f}\{-e_E + (r - y_E)\}, \quad (A.4)$$

$$\frac{1}{T_f}\sum_{i=1}^{I} \frac{dV_{E_i}}{dt} = -\frac{e_E}{T_f}(r - y_E). \quad (A.5)$$

Then, (A.4) and (A.5) yield

$$\frac{d\tilde{V}}{dt} = -\frac{e_E^2}{T_f} \leq 0. \quad (A.6)$$

Based on (A.1), (A.2), and (A.6), $\tilde{V}$ is positive and $d\tilde{V}/dt$ is negative except for the case of $e_E = 0$. From Lyapunov stability theory, $e_E \to 0$ as $t \to \infty$.

Next, we show that the tracking error $e \to 0$ as time $t \to \infty$. The derivative of $\varphi_{P_j}$ for a power-type battery $P_j$ ($j = 1, \cdots, J$) is given by

$$\frac{d\varphi_{P_j}}{dt} = K_{P_j}e = K_{P_j}(r - y) = K_{P_j}(r_P - y_P), \quad (A.7)$$

where $y = y_E + y_P$, $r_p = r - y_E$. Then, the derivative of $u_{P_j}$ is given by

$$\frac{du_{P_j}}{dt} = \frac{d\sigma_{P_j}(\varphi_{P_j})}{dt} = \bar{K}_{\sigma P_j}\frac{d\varphi_{P_j}}{dt}, \quad (A.8)$$

where

$$\bar{K}_{\sigma P_j} = \begin{cases} K_{\sigma P_j} & \varphi_{P_j}(\varphi_{P_j} - \varphi_{P_j}^{\text{Limit}}) < 0 \\ 0 & \text{otherwise} \end{cases}, \quad (A.9)$$

$\varphi_{P_j}^{\text{Limit}} = \bar{\varphi}_{P_j}^D$ for $r_{P_j} > 0$, else $\varphi_{P_j}^{\text{Limit}} = \underline{\varphi}_{P_j}^C$. $K_{\sigma P_j}$ is the slope in Fig. 3(b). Substituting (A.7) into (A.8), we have

$$\frac{du_{P_j}}{dt} = \bar{K}_{\sigma P_j}K_{P_j}(r_P - y_P), \quad (A.10)$$

and then

$$\frac{dy_P}{dt} = \sum_{j=1}^{J}\frac{du_{P_j}}{dt} = \left(\sum_{j=1}^{J}\bar{K}_{\sigma P_j}K_{P_j}\right)(r_P - y_P). \quad (A.11)$$

Given a feasible demanded power, some batteries operate with $\bar{K}_{\sigma P_j} = K_{\sigma P_j}$. As a result, we obtain $y_P \to r_P$ as $t \to \infty$ in (A.11). Hence,

$$\lim_{t \to \infty} e = 0 \quad (A.12)$$

holds. This completes the proof for Theorem 1.

## Appendix B. CONVENTIONAL CENTRALIZED CONTROL METHOD

The reference electric power for each battery depends on its own SOC. The reference electric power $r_{E_i}$ for energy-type battery $E_i$ is defined as

$$r_{E_i} = \frac{\Delta s_{E_i}}{\Delta s_{E_1} + \Delta s_{E_2} + \Delta s_{E_3} + \Delta s_{E_4} + \Delta s_{E_5}}r_E. \quad (B.1)$$

Here, $r_E$ satisfies

$$T_f \frac{dr_E}{dt} + r_E = r. \quad (B.2)$$

For $r_E > 0$ and $r_E < 0$, $\Delta s_{E_i} = s_{E_i} - s_{min}$ and $\Delta s_{E_i} = s_{max} - s_{E_i}$, respectively. Equation (B.1) means that the higher the battery SOC, the more the battery is discharged for $r_E > 0$. Conversely, the lower the battery SOC, the more the battery is charged for $r_E < 0$. The electric power $u_{E_i}$ for energy-type battery $E_i$ is expressed by

$$u_{E_i} = \begin{cases} \bar{U}_{E_i}^D & (r_{E_i} \geq \bar{U}_{E_i}^D) \\ r_{E_i} & (\bar{U}_{E_i}^D > r_{E_i} \geq \underline{U}_{E_i}^C) \\ \underline{U}_{E_i}^C & (\underline{U}_{E_i}^C \geq r_{E_i}) \end{cases}. \quad (B.3)$$

Similarly, the reference electric power $r_{P_j}$ for power-type battery $P_j$ is defined as

$$r_{P_j} = \frac{\Delta s_{P_j}}{\Delta s_{P_1} + \Delta s_{P_2} + \Delta s_{P_3} + \Delta s_{P_4} + \Delta s_{P_5}}(r - r_E). \quad (B.4)$$

Here, for $r > r_E$ and $r < r_E$, $\Delta s_{P_j} = s_{P_j} - s_{min}$ and $\Delta s_{P_j} = s_{max} - s_{P_j}$, respectively. The electric power $u_{P_j}$ for power-type battery $P_j$ is expressed by replacing the subscript $E_i$ in (B.3) with $P_i$.